\documentclass[10pt,journal,final,twocolumn]{IEEEtran}
\usepackage{amsmath,amsfonts}

\usepackage{algorithm}
\usepackage{array}
\usepackage{textcomp}
\usepackage{stfloats}
\usepackage{url}
\usepackage{verbatim}
\usepackage{graphicx}
\usepackage{cite}

\usepackage{mathrsfs}
\usepackage{amsthm}
\usepackage{amssymb}
\usepackage{latexsym}
\usepackage{subfigure}
\usepackage{pifont}
\usepackage{enumerate}
\usepackage{bm}
\usepackage{makecell}
\usepackage{algpseudocode}
\usepackage{dirtytalk}
\usepackage{tabularx}
\usepackage{mathtools}
\usepackage{booktabs}
\usepackage{hyperref}
\DeclareFontFamily{U}{wncy}{}
    \DeclareFontShape{U}{wncy}{m}{n}{<->wncyr10}{}
    \DeclareSymbolFont{mcy}{U}{wncy}{m}{n}
    \DeclareMathSymbol{\Sh}{\mathord}{mcy}{"58}

\begin{document}

\title{GBSense: A GHz-Bandwidth Compressed Spectrum Sensing System}

\author{
Zihang Song, Xingjian Zhang, Zhe Chen,
Rahim Tafazolli and
Yue Gao
\thanks{This work was supported by the Engineering and Physical Sciences Research Council of United Kingdom under the Grant EP/R00711X/2.}
\thanks{Zihang Song is with the Department of Engineering, King's College London, Strand, London, WC2R 2LS, United Kingdom (email: zihang.song@kcl.ac.uk).}
\thanks{Xingjian Zhang is with Guangdong Provincial Key Laboratory of ACNT, Harbin Institute of Technology (Shenzhen), Shenzhen, China, (email: x.zhang@hit.edu.cn)}
\thanks{Zhe Chen and Yue Gao are with the School of Computer Science, Fudan University, Shanghai 200433, China (e-mail: zhechen@fudan.edu.cn, yue.gao@ieee.org).}
\thanks{Rahim Tafazolli is with 5GIC \& 6GIC, Institute for Communication Systems (ICS), University of Surrey, Guildford, UK (email: r.tafazolli@surrey.ac.uk)}
}


\maketitle

\begin{abstract}
This paper presents GBSense, an innovative compressed spectrum sensing system designed for GHz-bandwidth signals in dynamic spectrum access (DSA) applications. GBSense introduces an efficient approach to periodic nonuniform sampling, capturing wideband signals using significantly lower sampling rates compared to traditional Nyquist sampling. By integrating time-interleaved analog-to-digital conversion, GBSense overcomes the hardware complexity typically associated with traditional multicoset sampling, providing precise, real-time adjustable sampling patterns without the need for analog delay circuits. 
The system's ability to process signals with a 2 GHz radio frequency bandwidth using only a 400 MHz average sampling rate enables more efficient spectrum monitoring and access in wideband cognitive radios. Lab tests demonstrate 100\% accurate spectrum detection when the spectrum occupancy is below 100 MHz and over 80\% accuracy for occupancy up to 200 MHz. Additionally, an integrated system utilizing a low-power Raspberry Pi processor achieves a low processing latency of around 30 ms per frame, demonstrating the system's potential for real-time applications in cognitive radio networks, 5G, and future 6G infrastructures. 
\end{abstract}


\section{Introduction}\label{se_introduction}
\IEEEPARstart{E}{fficient} spectrum utilization has become urgent with the developments of 5G and forthcoming 6G technologies \cite{ITU2020, 3GPP2018, zhang2019_6g, song2021survey}. This urgency is accentuated in the burgeoning Internet of Things (IoT) network, where a large number of devices compete for limited bandwidth. In recent years, dynamic spectrum access (DSA) has been considered a promising technology to address the increasing demand for spectrum resources. It allows the unlicensed secondary users to opportunistically utilize the frequency band licensed to primary users, thus significantly improving the spectrum utilization efficiency \cite{cr1999, Haykin2005, zhao2007survey}.

The effective implementation of DSA relies on the ability to accurately and efficiently sense the spectrum. Traditional spectrum sensing techniques are predominantly based on the Nyquist sampling theorem, which stipulates that a signal must be sampled at least twice its highest frequency to ensure accurate reconstruction. This constraint poses significant challenges to the deployment of DSA in scenarios involving wideband signals, where the required Nyquist rate can be prohibitively high, leading to the need for power-intensive and expensive high-speed analog-to-digital converters (ADCs) and digital signal processing units \cite{tian2008_compressive}.

To mitigate these challenges, compressed spectrum sensing (CSS) has been introduced as an innovative solution \cite{tian2007cs,song2022approaching}. CSS leverages the sparse nature of wideband signals in the frequency domain, allowing for sub-Nyquist sampling without compromising the ability to accurately reconstruct the original signal using compressed sensing (CS) recovery algorithms \cite{candes2005decoding}. CSS has the potential to significantly reduce the computational and resource burdens associated with spectrum sensing in DSA applications \cite{song2022approaching}. However the practical deployment of CSS in large-scale networks remains challenging due to hardware limitations, particularly in achieving precise timing control for sub-Nyquist sampling and managing the high data throughput required for real-time applications. The design and implementation of CSS systems that can operate efficiently in real-world network environments is therefore a crucial area of research.

In this paper, we present GBSense, a novel hardware implementation of a CSS system tailored for wideband, real-time DSA in modern wireless networks. GBSense provides a hardware-friendly, energy-efficient solution that can be integrated into cognitive radio networks, 5G, 6G, and IoT environments. The primary contributions of this work are as follows:

\subsection{Main Contributions} 
\begin{enumerate} \item The GBSense system leverages the\textit{ time-interleaved analog-to-digital conversion }(TI-ADC) technique, providing an efficient and easily implementable \textit{periodic non-uniform sampling} (PNS) solution. This addresses the challenges posed by analog delay circuits in traditional \textit{multi-coset sampling} (MCS) setups by offering precise, real-time adjustable delay control without the need for complex analog components. \item The GBSense system reduces the data transmission burden between the sampler and the processor, allowing for the use of low-power processors to achieve near real-time signal processing, thus improving overall system efficiency and reducing power consumption. \item We demonstrate that the GBSense system can process signals with a 2 GHz radio frequency (RF) bandwidth using only a 400 MHz average sampling rate. Lab tests show that the system can accurately reconstruct the signal spectrum with 100\% confidence when the spectrum occupancy is below 100 MHz and over 80\% confidence for spectrum occupancy up to 200 MHz. \item An integrated CSS system built around the GBSense core sampler boards and a low-power Raspberry Pi processor has been developed, achieving a low processing latency of around 30 ms for a frame, demonstrating strong real-time performance. \end{enumerate}

The rest of this paper is organized as follows: Section II provides a brief review of existing architectures and hardware realizations of CSS systems. Section III presents the preliminaries necessary for understanding CSS systems, with a focus on the challenges of implementing the traditional MCS architecture in hardware. Section IV offers an overview of the core principles behind the GBSense system, highlighting how time-interleaved sampling and compressed sensing enable sub-Nyquist spectrum sensing. Section V outlines the system architecture and hardware design. Section VI discusses the performance validation, including testing and real-world implementation results. Section VII concludes the paper.

\section{Related Work}
The fundamental aim of sampling at a sub-Nyquist rate in CSS is to efficiently project a sparse signal into a lower-dimensional space through a linear transformation. CSS sub-Nyquist sampling strategies can be categorized into two primary groups:
\subsubsection{Pseudo-random Mixing (PRM)}
PRM-based architectures achieve signal dispersion in the analog domain by mixing the input signal with analog pseudo-random sequences (PRSs), followed by sub-Nyquist-rate uniform sampling. This mixing process corresponds to a convolution in the frequency domain. A notable example is the random demodulator (RD) \cite{Tropp2010pd}, which uses fast PRSs for mixing, followed by low-rate integration and sampling \cite{kirolosAnalogtoInformationConversionRandom2006a}. PRM-based architectures, such as the \textit{Random-Modulation Pre-Integrator} (RMPI) and \textit{Spread Spectrum Random Modulator Pre-Integrator} (SRMPI), offer a way to achieve sub-Nyquist sampling by adding pseudo-random modulation steps. However, its demand for full Nyquist-rate modulation introduces inefficiencies in power consumption and increases complexity \cite{yoo2012rmpi,mamaghanian2012srmip}. The modulated wideband converter (MWC), a multi-channel extension of the RD \cite{mishali2010theory}, enhances spectrum preservation through multi-channel mixing with diverse PRSs. \textit{Xampling}, based on the MWC architecture, similarly attempts to sample wideband signals at sub-Nyquist rates. However, the need for generating multiple high-speed PRSs and mixing in the analog domain without distortions continues to add complexity to the hardware \cite{mishali2011xampling, eldar2012compressed, baransky2014sub}.
More recent innovations, such as \textit{memristor-based compressed sensing}, have shown potential in reducing hardware complexity by leveraging the analog computing capabilities of memristors \cite{wang2023memristor}. While promising for high-speed applications, this technology suffers from the analog nature of memoristor devices related to programming incccuracy, thermal noise, variability and conductance drift \cite{song2024xpikeformer}. 

\subsubsection{Non-uniform Sampling (NUS)}
NUS-based architectures directly digitize the raw input signal non-uniformly at an average rate below the Nyquist rate. Fourier Transform is typically applied in the digital domain to obtain aliased frequency-domain measurements. Unlike PRM-based methods, NUS does not require the generation of PRSs or their mixing with the input signal, making it more efficient. For example, a custom NUS chip digitizes signals in the 800 MHz to 2 GHz band at an average rate of 236 MSps, using off-the-shelf ADCs and a sample-and-hold circuit \cite{candes2012nuschip}. However, this system drives ADC with an irregular clock which casues high complexity and cost to generate. Also, the irregular clock would disrupt synchronization, increase jitter, and complicate downstream processing \cite{davenport2012signal}. \textit{Direct Under-Sampling Compressed Sensing} (DUS-CS) faces similar issues with jitter and time precision \cite{sun2019duscs}. The \textit{Non-Uniform Wavelet Band-pass Sampling} (NUWBS) method leverages wavelet transforms for spectrum sensing but is hindered by complex signal reconstruction and control requirements \cite{daponte2018nuwbs}.

A specific form of NUS is PNS, where samples are taken at varying intervals in a repeating pattern. The multi-coset sampler (MCS) \cite{venkataramani2001optimal} is an architecture designed to implement PNS by applying different time delays across parallel low-rate sampling channels. This allows the acquisition of non-uniform samples without need to generate non-uniform clocks\cite{celebi2013multi}. As a result, most studies on MCS-based CSS remain at the theoretical and simulation levels \cite{cohen2014sub, tropp2010beyond}. Notably, \textit{Multicoset Sampling based on Compressed Sensing} (CSMC) attempted to implement traditional MCS using analog time-delay circuits \cite{jingchao2015csmc}. However, the need for highly precise time delays introduces significant challenges in the design of delay circuits, particularly regarding accurate delay control, constant frequency response, and flexibility \cite{jingchao2015csmc}. These are precisely the issues we aim to address in our GBSense implementation.

\section{Preliminaries}\label{section_math}
\subsection{Periodic Nonuniform Sampling}
\begin{figure}[t]
    \centering
    \includegraphics[width=8.5cm]{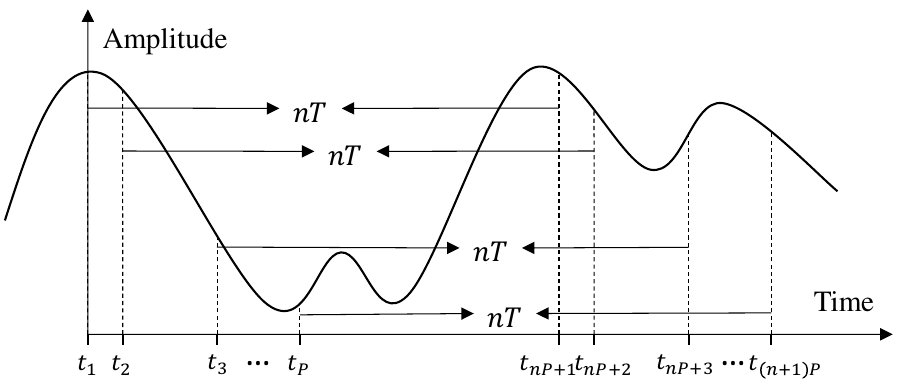}
    \caption{A simple illustration of periodic non-uniform sampling, where $n \in\mathbb{Z}^{+}$}
    \label{fig:pns}
\end{figure}

PNS is a sampling strategy where the time interval between successive sample points is not constant but follows a periodic pattern. A PNS sampling scheme can be characterized by the following parameters:
\begin{enumerate}
    \item Pattern period $T$: the temporal length of the interval during which the nonuniform pattern recurs.
    \item Relative offsets $\mathcal{T}_P$: 
    \begin{equation}
        \mathcal{T}_P=\{t_p|p=1,\ldots,P, 0=t_1<t_2<\ldots<t_p<T\},
    \end{equation}
    where $P$ denotes the number of samples in each pattern period.
\end{enumerate}
As depicted in Figure \ref{fig:pns}, each set of $P$ successive samples exhibit a non-uniform pattern. Despite this non-uniformity within each group, the time interval $T$ between the $i$-th sample and the $(i+P)$th sample remains constant.

\subsection{Multi-coset Sampling Framework}
\begin{figure}[t]
\centering
\hspace{-3mm}\subfigure[]{\includegraphics[height = 4.5cm]{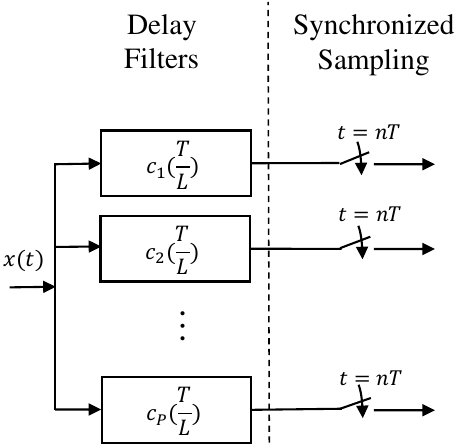}\label{fig:mcs_traditional}}
\subfigure[]{\includegraphics[height = 4.5cm]{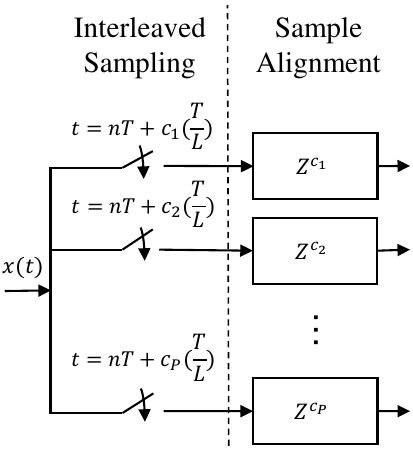}\label{fig:mcs_gbsense}}
\caption{(a) The delay-and-sampling architecture assumed by the traditional MCS and (b) the time-interleaved sampler adopted by the GBSense system.}
\end{figure}

MCS offers a feasible solution by achieving PNS through multiple \textit{cosets} selected from a uniformly spaced grid with different offsets. A conceptual illustration of MCS is shown in Figure \ref{fig:mcs_traditional}, where $P$ ADCs, each operating at a constant sampling rate $f_{\mathrm{s}}=1/T$, process $P$ temporally shifted copies of the original analog signal. The temporal offsets are defined by the set $\mathcal{T}$, relative to a common reference point. Each ADC thereby captures a distinct subset of the expected PNS samples, and these subsets are interrelated through known time shifts. 

The relative offsets $\mathcal{T}$ can be represented discretely by defining an integer $L>P$ and a \textit{sampling pattern}:
\begin{equation}\mathcal{C}_{\{P,L\}}=\{c_p\in\mathbb{Z}|p=1,\ldots,P, 0=c_1<c_2<\ldots<c_p<L\}.\end{equation} 
satisfying $\frac{c_p}{L}T=t_p$, where $L$ represents the granularity (or the resolution) of temporal offset. For analog input $x(t)$, the $n$-th sample at the $p$-th MCS channel is formulated as
\begin{equation}
    x_{c_p}[n] = x\left(\frac{c_p}{L}T+nT\right), \quad p=1,2,...,P.
\end{equation}

\subsection{Spectrum Reconstruction with MCS Samples}
The MCS system has an average sampling rate of \(Pf_{\mathrm{s}}\). Based on the Whittaker-Shannon-Nyquist sampling theorem, this rate is insufficient to reconstruct a signal with a bandwidth of \(Lf_{\mathrm{s}}\). However, it has been shown that if there are only \(K\) active bands, each with a bandwidth less than \(f_{\mathrm{s}}\), the signal can be uniquely reconstructed from the MCS samples when \(K \leq \frac{P}{2}\) \cite{song2022approaching}. This unique reconstruction is typically achieved by solving the following optimization problem \cite{song2023numerical}:
\begin{equation}
\mathop{\arg\min}_{\mathbf{\Theta}} \| \mathbf{\Theta} \|_{2,0} \quad \text{subject to} \quad \| \mathbf{X} - \mathbf{A} \mathbf{\Theta} \|_{\text{F}} < \epsilon,
\end{equation}
where \(\mathbf{X}\) is a \(P\)-row matrix, with each row representing the Fourier transform of the samples acquired in each MCS lane, \(\mathbf{\Theta}\) denotes the original signal spectrum, \(\mathbf{A}\) is the \textit{sensing matrix} determined solely by \(L\) and \(\mathcal{C}_{\{P,L\}}\), and \(\epsilon\) is a small value accounting for error. The detailed formulation of this problem is given in Appendix A.

\subsection{Challenges in MCS Hardware Realization}\label{se:MCS_challenge}
The MCS framework represents a promising approach to PNS. However, translating these concepts into practical hardware implementations poses several technical challenges:

\subsubsection{Analog Delay Imperfection} The MCS architecture relies on precise analog delays to generate non-uniform sampling patterns, typically achieved through all-pass filters. Designing these filters to maintain constant magnitude and linear phase response is complex and costly \cite{papoulis1977generalized, delorimier2004allpass}. Achieving precise delay control is particularly challenging as these filters are prone to noise and circuit imperfections, resulting in inaccuracies that complicate the implementation of controlled analog delays \cite{anderson2007delay,marques2002high}.

\subsubsection{Lack of Flexibility} CSS systems require dynamically adjustable sampling patterns to adapt to varying signal characteristics. However, the fixed delay properties of analog all-pass filter circuits restrict their flexibility, limiting their adaptability to different signal environments. This lack of dynamic adjustment hampers their application in diverse real-world scenarios \cite{song2024nonuniform}.

\subsubsection{Throughput and Processing Demands} MCS implementations often necessitate parallel data streaming and the simultaneous execution of fast Fourier transforms (FFT) across multiple channels to maintain real-time performance. This results in significant demands on throughput and processing capabilities, particularly for low-power back-end controllers, which are critical in many practical applications.

\section{GBSense System Design}
To address the challenges mentioned in Section \ref{se:MCS_challenge}, we introduce the GBSense system, a specialized hardware implementation of the PNS framework introduced in Section \ref{section_math}. GBSense not only addresses the issues of sampling accuracy and adjustability but also significantly reduces processing capability requirements, utilizing readily available electronic components.
\subsection{Time-interleaved Sampling}
The sampling architecture of the GBSense system is inspired by the TI-ADC technique \cite{vogel2006time}. Unlike the traditional MCS approach, which relies on analog delays and phase-synchronized ADC sampling, the GBSense system directly digitizes multiple copies of the signal $x(t)$ without introducing any offsets to the signal itself. Instead, the $P$ sampling clocks are individually adjusted to a unique offset of $\frac{c_p}{L}T$ relative to a common reference to achieve non-uniform sampling, as illustrated in Fig. \ref{fig:mcs_gbsense}. Precise phase control of these clocks is achieved through a clock distribution circuit based on phase-locked loops (PLLs), as will be detailed in Section \ref{se:hardware}. 

For mathematical rigor, Fig. \ref{fig:mcs_gbsense} also conceptually shows a digital right shift of \(c_p\) cycles of \((T/L)\) to realign the \(P\) sample sequences. However, we later demonstrate that this realignment does not require an explicit digital unit driven by a high-speed clock of frequency \(L/T\). Instead, all \(P\) data streams are inherently aligned by the standardized serial interface within the input buffer of the receiver logic device (LD), which operates at a lower clock rate of \(1/T\). A mathematical proof of the equivalence between Fig. \ref{fig:mcs_traditional} and Fig. \ref{fig:mcs_gbsense} is provided in Appendix B.

The time-interleaved sampling technique allows the GBSense system to bypass the technical challenges associated with analog delay circuits of the traditional MCS architecture. Key advantages include (1) \emph{Flat Frequency Response}: Each lane's signal maintains its integrity because it only passes through transmission lines, enabling high-bandwidth frequency response with minimal signal distortion; (2) \emph{Precision}: Precise offset control of sampling clocks can be achieved using commercially available PLL chips with ultra-low jitter and skew rates (down to femtosecond and sub-picosecond levels), significantly enhancing sample fidelity; (3) \emph{Flexibility}: The sampling pattern can be easily adjusted by apply offset to the sampling clocks through digital configurations to the PLL chip's control registers.

\subsection{Data Synchronization}\label{se:jesd}
A significant challenge introduced by time-interleaved sampling is achieving precise data realignment across $P$ ADCs at the receiver LD input buffer. The non-aligned digitization instances, combined with nondeterministic link latency arising from factors like device variations and differences in lane lengths, can lead to misalignment of data between channels. To address these issues, the GBSense system leverages the JESD204B high-speed serial interface \cite{JESD204B} to ensure reliable data realignment. 

\begin{figure}[t]
\centering\includegraphics[width = 9cm]{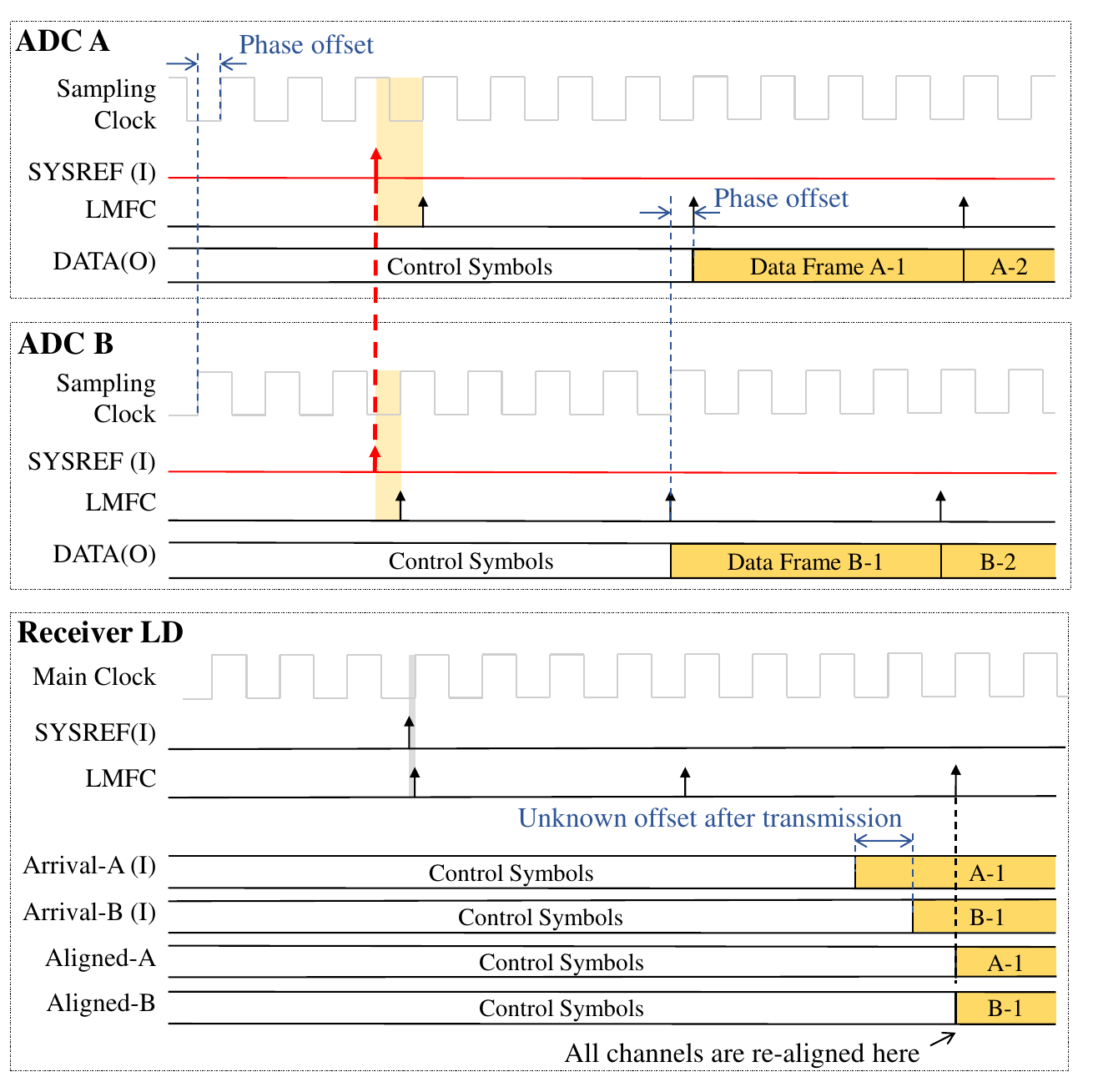}
\caption{A simplified illustration of the data synchronization process, facilitated through the JESD204B protocol. The terms (I) and (O) respectively denote input and output signals. Both the ADCs and the receiver LD receive a paired sampling clock and SYSREF signal from the same clock distribution PLL, synchronously initiating the internal LMFC. The ADCs begin sampling and transmitting valid data at the start of the next LMFC cycle. Due to lane latency uncertainties, the sampled data reach the LD at unpredictable intervals. However, these data are buffered and subsequently realigned at the receiver LD's next valid LMFC to ensure proper synchronization.}
\label{fig:jesd}
\vspace{-5mm}
\end{figure}

Fig. \ref{fig:jesd} illustrates a simplified view of the synchronization process, focusing on two ADC channels and a single receiver LD for clarity. Prior to system initialization, the ADCs only transmit control symbols instead of real samples. The synchronization procedure begins with the clock distribution circuit sending a highly synchronized 'SYSREF' pulse to each ADC. Upon receiving the SYSREF pulse, each ADC resets its \textit{Local Multi-frame Clock} (LMFC) to the next rising edge of its sampling clock. The ADCs then begin transmitting real digitized samples, packed into data frames, at the next LMFC rising edge. The length of each data frame is typically set to include multiple samples (e.g., 32, 64), resulting in a significantly slower LMFC compared to the sampling clock. The LD subsequently aligns the initial data frames from all ADCs to the first LD LMFC after detecting real signal samples from the slowest lane. This extended LMFC period enhances the timing margin, simplifying the alignment sequencing process and ensuring that data from each ADC reaches the receiver in perfect synchrony.

Once the connections are established, the JESD204B interface continuously monitors the timing alignment of data from each ADC. If any discrepancies are detected, the system automatically restarts the initialization process to realign the LMFCs. This ongoing synchronization ensures that the non-uniform sampling of the GBSense system remains consistent across all ADCs, regardless of variations in individual clock timings. 

\subsection{Data Decimation}\label{se:decimation}
Reconstructing the signal spectrum from PNS samples requires executing CS recovery algorithms. For a CSS task, we consider a speed-optimized combination of algorithms: Fast Fourier Transform (FFT) for generating \(\mathbf{X}\) with complexity \(O(N \log N)\), Sparse Orthogonal Matching Pursuit (SOMP) for spectrum reconstruction \(\mathbf{\Theta}\) \cite{tropp2005somp} with complexity \(O(N)\), and Energy Detection (ED) for identifying active channels \cite{cr1999} with complexity \(O(N)\). The computational load is reduced by processing smaller windows, \(N\), though this increases aliasing risk in the FFT output. 

To address this, we apply \textit{data decimation} and the following parallel FFT in the LD to lower computational costs without sacrificing accuracy \cite{song2019real}. Let \(d\) be the decimation factor. The \(N\)-point time-domain data \(x_{c_p}[n],\ n = 0, 1, \ldots, N-1\), is divided into \(d\) equal segments (assuming \(N\) is divisible by \(d\)). By overlapping and summing these segments, the data is compressed to \((N/d)\)-point data per lane, reducing the processing load without introducing additional aliasing.

\section{Hardware Implementation}\label{se:hardware}
\begin{figure*}
\centering  
\includegraphics[width=16cm]{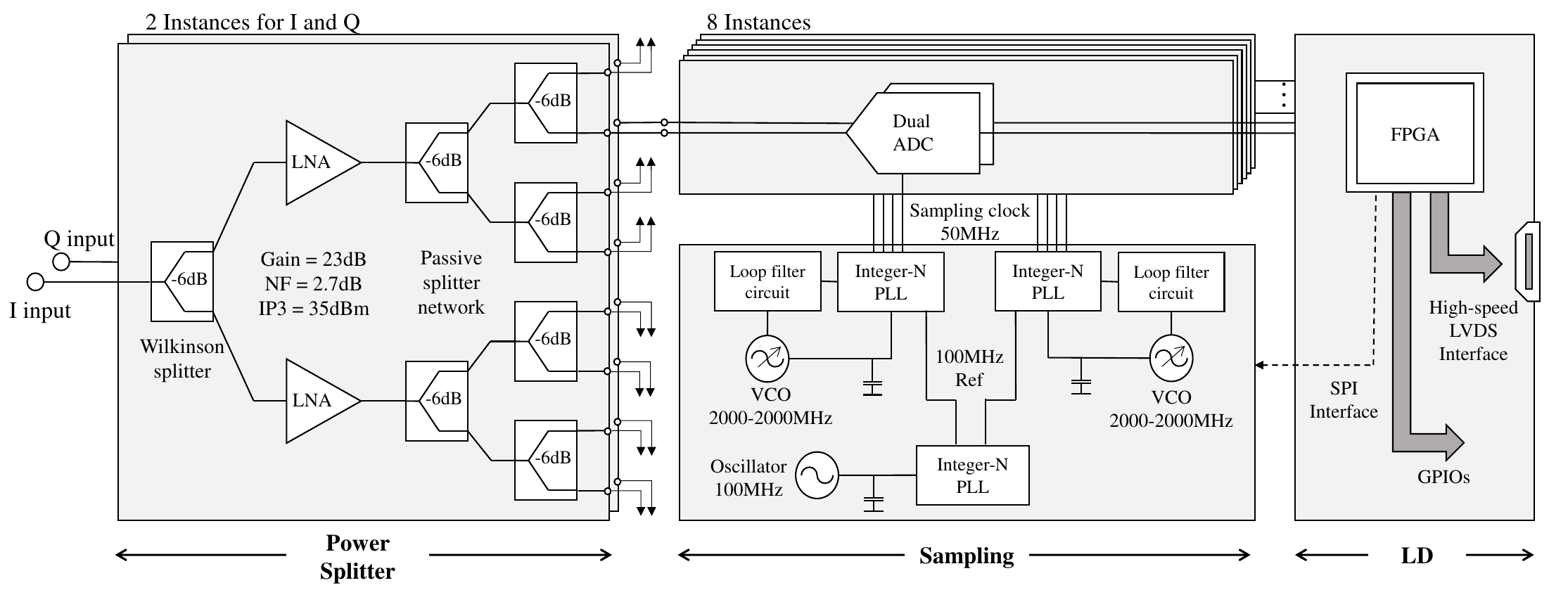}
\caption{An overall diagram of the circuit design of the multicoset sampling board.}
\label{fig:sysdiag}
\end{figure*}

The hardware implementation consists of three subsystems: a 1-to-8 power splitter, a PNS sampling unit, and an LD module. Each subsystem is implemented on separate circuit boards to minimize electromagnetic interference. A schematic diagram of the complete system is shown in Fig. \ref{fig:sysdiag}. The system accepts in-phase and quadrature (I/Q) component input signals with a 1 GHz single-sided (or 2 GHz double-sided) bandwidth.

\subsection{Power Splitter Subsystem}
The power splitter employs two identical power-splitting networks to separately divide the I/Q components into eight portions. Each network is structured into three layers: the first layer features a dual-matched amplifier for signal pre-amplification and division, which reduces noise and interference, and includes a monolithic microwave integrated circuit chip that provides an in-band gain of 20 to 25.1 dB, meeting the side-sense stationary processing requirements. The subsequent layers utilize a passive Wilkinson resistive splitter network with a Wye three-port configuration, resulting in a typical 6.02 dB loss between any two ports. Each splitter incorporates high-precision resistors ($\pm 0.1 \%$) to ensure uniform signal division. The two power splitter networks are mirrored on opposite sides of a PCB board, accommodating both single-ended and differential inputs and controlled by an RF switch chip for device compatibility. Edge-mounted connectors are used for input and output interfaces, maintaining high symmetry on the compact board.

\subsection{Time-interleaved Sampling Subsystem}
The sampling subboard, which receives $P$ pairs of signal copies from the power splitter, performs time-interleaved sampling. To mitigate high-frequency spurious interference from the switch power supply that could affect the dedicated interleaved sampling clocks, the design utilizes a cable connection to the power splitter instead of integrating them onto a single board.

The ADC array consists of eight dual-channel ADC devices (AD9250), where each I/Q component pair is digitized by one device using two embedded converters controlled by a shared input sampling clock. The clock distribution circuit manages the sampling pattern across the ADCs. 

A two-layer PLL network is deployed to generate the necessary clock and SYSREF signals. The first layer employs an open LTC6952 integer-N PLL (primary) with an ultra-low phase noise oscillator, providing a 100 MHz input signal. The second layer consists of two integer-N PLL circuits (secondary), each driven by ultra-low phase noise voltage-controlled oscillators (VCOs) with a fixed output frequency of 2000 MHz. The output clock signals from the secondary PLLs are set to 50 MHz, achieved by dividing the 2000 MHz VCO signal by a factor of 40. The primary PLL supplies two synchronized 100 MHz reference clock signals to align the internal clocks of the secondary PLLs. Additionally, two SYSREF signals are provided alongside the reference clocks, enabling synchronized control of the secondary SYSREF outputs. Both the reference clock lanes and SYSREF lanes are meticulously designed to be of identical length, ensuring highly synchronized signals at the secondary PLLs. The output clock signals from each secondary PLL are distributed as the sampling clocks for 4 of the 8 ADC devices, respectively. All 8 physical lanes for the sampling clocks are also designed to be of equal length. Moreover, one of the secondary PLLs also provides a synchronized clock to the LD, thereby ensuring coherent timing across the entire system. 

The system enables independent phase configuration of the output clock signals, with a digitally controlled phase delay providing a minimum step size of half the VCO period, given by:
\begin{equation}\label{eq_step}
    \Delta t_{\min} = \frac{1}{2} \times \frac{1}{2000\; \text{MHz}} = 250\; \text{ps},
\end{equation}
which corresponds to 1/80 of a cycle of the 50 MHz sampling clock. Consequently, \(L\) can be any integer divisor of 80, such as 80, 40, or 20. A carefully designed power supply network delivers low-ripple direct current (DC) power levels. Both 2000 MHz VCOs are powered by a common DC source, ensuring identical peak-to-peak voltages. This synchronization guarantees that the rising edge of the clock signal consistently crosses the trigger level at the same offset, reducing phase errors. 

\subsection{Logic Device Subsystem}
The LD subsystem is designed to preprocess the data streams from the ADC array and facilitate communication with general-purpose computing devices. A Lattice ECP3-150EA FPGA device implements the JESD204B receiver functionality at its input buffer, re-aligning the sample sequences from the ADCs and performing data decimation and FFT operations. The FPGA also includes a control interface that allows real-time configuration and adjustment of the PLL and ADC devices. The board is equipped with high-speed LVDS interfaces for efficient communication with energy-intensive controllers, as well as general-purpose input/output (GPIO) connectors for low-speed communication with compact computing platforms.

\begin{figure}[tb]
    \centering
    \includegraphics[width=1\linewidth]{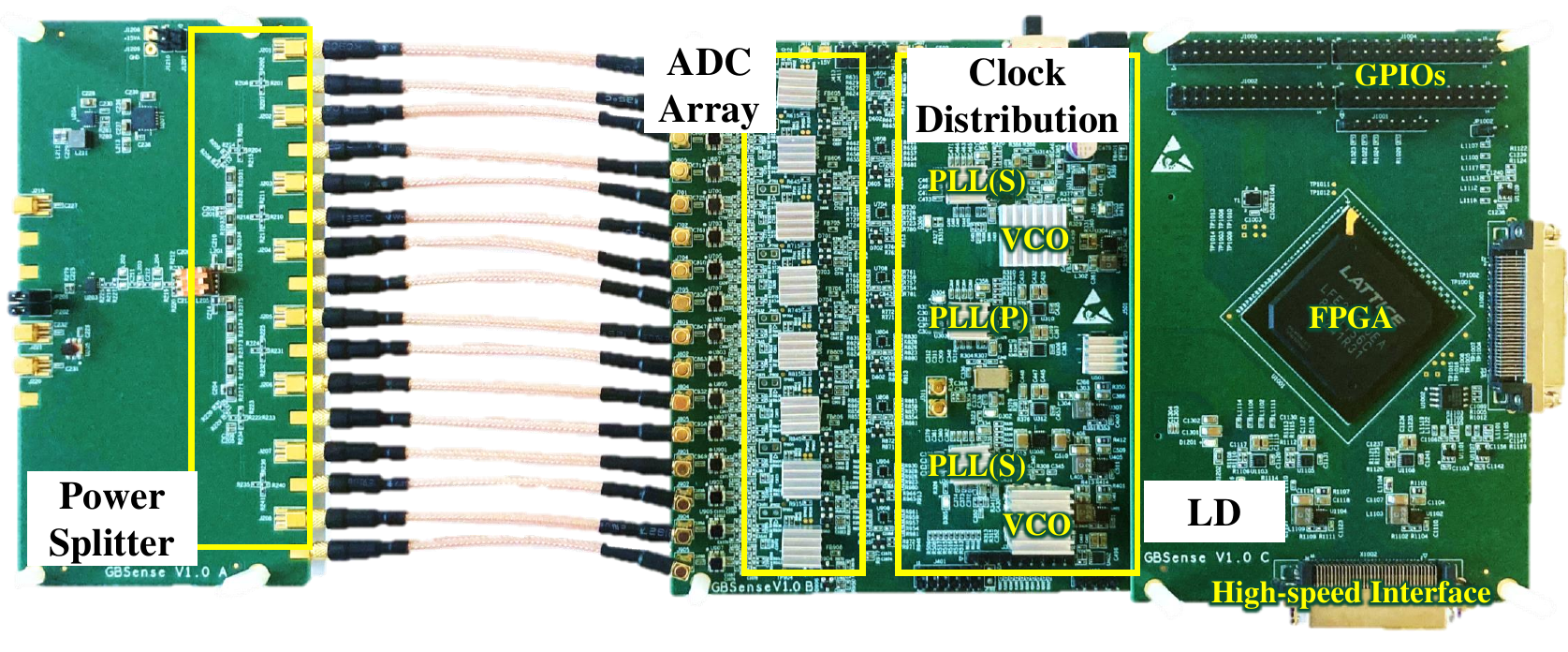}
    \caption{A photo of the GBSense circuit boards.}
    \label{fig_board}
\end{figure}

The photo of the GBSense circuit board is shown in Fig. \ref{fig_board}, where the power splitter board is connected to the sampling board via a coaxial cable. The specifications of the GBSense system parameters are provided in Table \ref{tb:sepcifications}.

\begin{table}[ht]\label{tb:sepcifications}
\centering
\caption{Specifications of GBSense system parameters}
\begin{tabular}{cc}
\hline
\textbf{Parameter}           & \textbf{Value} \\ \hline
baseband I/Q bandwidth        & $1$ GHz       \\
RF bandwidth        & $2$ GHz           \\
number of physical lanes $P$   & 8              \\
sampling rate/lane $f_{\text{s}}$        & 50 MSps         \\
average sampling rate $Pf_{\text{s}}$       & 400 MSps        \\
minimum clock delay resolution  & 250 picosecond           \\
window length $N$              & 256/512/1024 pts        \\ \hline
\end{tabular}
\end{table}

\section{System Validation and Testing}\label{se_simulation}

\subsection{Electronic Features}
The critical circuits, namely the power splitter and the clock distribution, are first individually tested. The power splitter board has been tested over a range from 0.01 to 1 GHz, exhibiting a peak gain of 6 dB at lower frequencies that decreases to around 1.7 dB at 1 GHz. An in-band spectrum ripple, centered at 505 MHz, shows an approximate 2.2 dB deviation within a $\pm 495$ MHz band. Isolation characteristics reveal about -7 dB in-band isolation with adjacent ports sharing a common 3rd-layer Wye splitter, around -20 dB with ports sharing a 2nd-layer splitter, and less than -30 dB in-band isolation with four ports connected to the other output of the dual-matched amplifier. 

\begin{figure}[t]
    \centering
    \includegraphics[width=9cm]{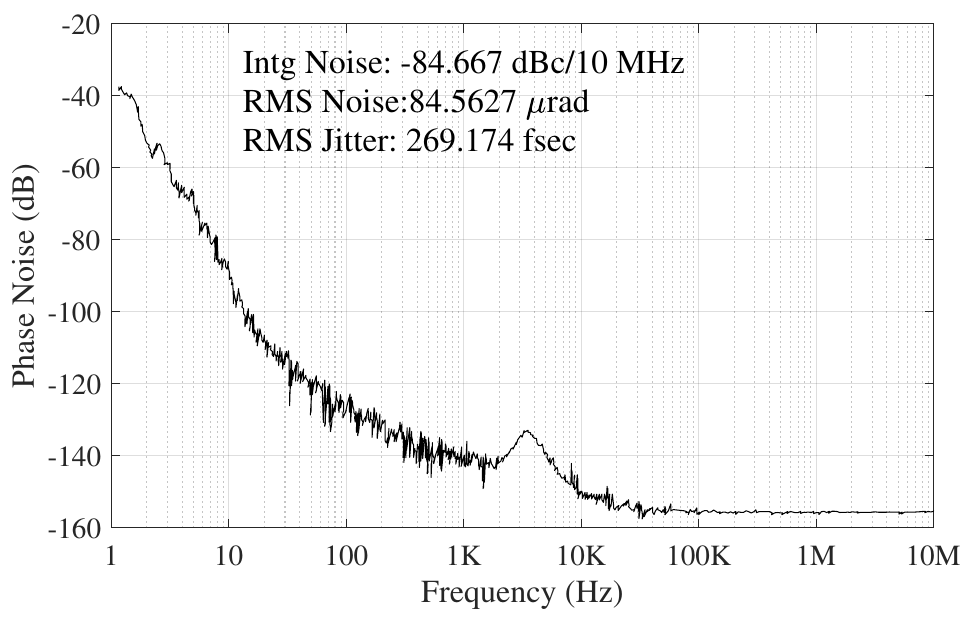}
    \caption{Phase Noise of the 50 MHz sampling clock generated by the clock distribution circuit, measured by Keysight E5052B signal source analyzer.}
    \label{fig:phasenoise}
\end{figure}

\begin{table}[htbp]
    \centering
    \caption{Electrical Specifications of the Power Splitter Board}
    \begin{tabular}{ll}
        \toprule
        \textbf{Parameter}  &  \textbf{Value} \\
        \midrule
        Frequency range & 10 MHz $\sim$ 1 GHz \\ 
        Gain & 1.7 $\sim$ 6.1 dB \\ 
        In-band spectrum ripple & 2.2 dB \\ 
        In-band output isolation & -38 $\sim$ -6.7 dB \\ 
        \bottomrule
    \end{tabular}
    \label{tab:pwsplt}
\end{table}

To test the clock distribution circuit, we routed the generated 50 MHz clock through a test port and measured the phase noise using the Keysight E5052B signal source analyzer. The maximum phase noise measurement, within the frequency range of 1 Hz to 10 MHz, is shown in Fig. \ref{fig:phasenoise}. The corresponding time-domain root mean square (RMS) jitter caused by the phase noise is 269.174 femtoseconds, which is about 0.0013\% of the 20 nanosecond sampling period. This jitter introduces approximately 0.11\% absolute error to the minimum phase offset step, as calculated in \eqref{eq_step}. Additionally, this level of jitter corresponds to a signal-to-noise ratio (SNR) ceiling of approximately 55 dB for full-scale sampling \cite{ad9250datasheet}.


\subsection{System Validation with Controlled Signal Inputs}

\begin{figure}[t]
\centering
\subfigure[]{\includegraphics[width=1\linewidth]{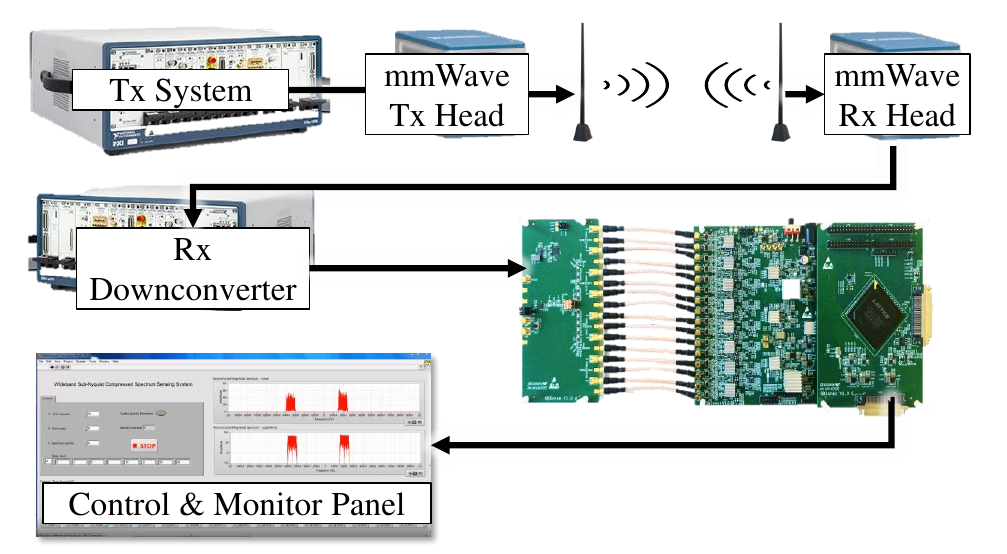}\label{fig_system}}
\subfigure[]{\includegraphics[width=1\linewidth]{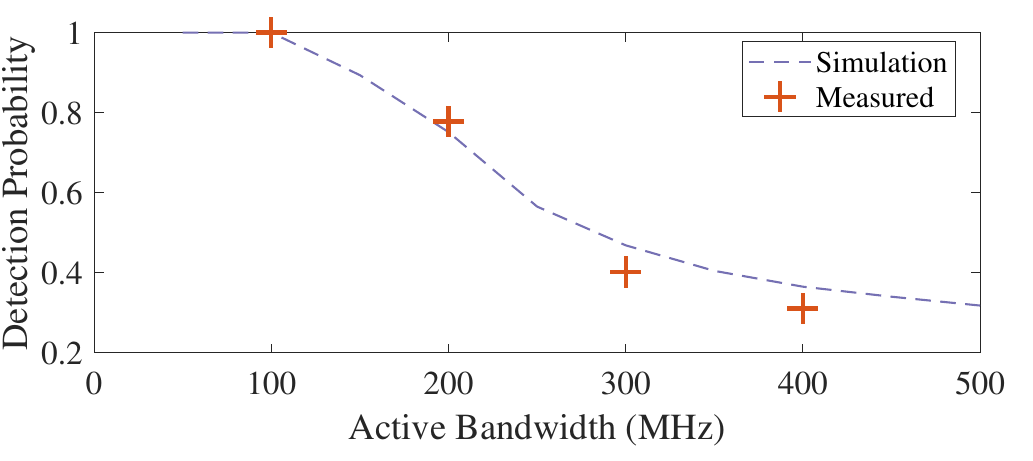}\label{fig_pd}}
\caption{(a) A general overview of the testbed of the GBSense boards. (b) Detection probability achieved by the testbed, compared with simulation results obtained under an SNR of 10 dB.}
\end{figure}

The overall sampler of GBSense system was validated using the National Instruments millimeter-wave transceiver system (MTS), which includes a modularly designed transmitter and receiver. The transmitter generates a complex baseband signal with a total bandwidth of 1 GHz for each I/Q component, occupying the baseband frequency range \([-1, 1]\) GHz. When modulated to the 28.5 GHz carrier, the resulting RF signal spans a total of 2 GHz bandwidth. This RF signal carries up to 8 groups of orthogonal frequency-division multiplexing subcarriers, each with a maximum bandwidth of 100 MHz, meaning that only up to 800 MHz of the 2 GHz RF bandwidth is active at any given time.

To test the GBSense system, we routed the downconverted I/Q components (each with a 1 GHz baseband bandwidth) from the receiver, bypassing the original sampling and preprocessing stages of the NI testbed. Instead, these I/Q signals were fed into the GBSense sampler for PNS. The samples processed by GBSense were then fed back into the receiver's CPU controller for spectrum reconstruction. In this way, the GBSense system replaced the original sampling and preprocessing components of the NI system. The pre-processed data \(\mathbf{X}\) was transmitted through the high-speed LVDS interface, and the spectrum reconstruction algorithm was executed on the testbed CPU. A user interface developed in LabVIEW displayed the reconstructed spectrum. The testbed implementation is illustrated in Fig. \ref{fig_system}.

We set \(L = 40\), meaning the offset between any two sampling clocks is an integer multiple of 500 ps, and the entire \([-1, 1]\) GHz spectrum is divided into 40 subbands of 50 MHz each. The observation window length is set to \(N = 1024\) points. The system is tested by generating 1 to 4 non-overlapping transmissions, each with a 100 MHz bandwidth at known central frequencies on the transmitter. The spectrum is reconstructed from the PNS samples using the SOMP algorithm on the receiver's CPU controller. The detection probability of the reconstructed spectrum, relative to the ground truth, is calculated as:
\begin{equation}
    \text{Detection Probability}=\frac{\# \text{Correctly detected active subbands} }{ \# \text{Active subbands}}.
\end{equation}
The results are compared with a reference curve simulated at 10 dB SNR, as shown in Fig. \ref{fig_pd}. As observed, the GBSense testbed results align well with the simulated expectations, achieving accurate reconstruction with 100\% confidence when the spectrum occupancy is below 100 MHz and over 80\% confidence for spectrum occupancy up to 200 MHz.

\subsection{System-level Demonstration}
\begin{figure}[t]
\centering
\subfigure[]{\includegraphics[width=1\linewidth]{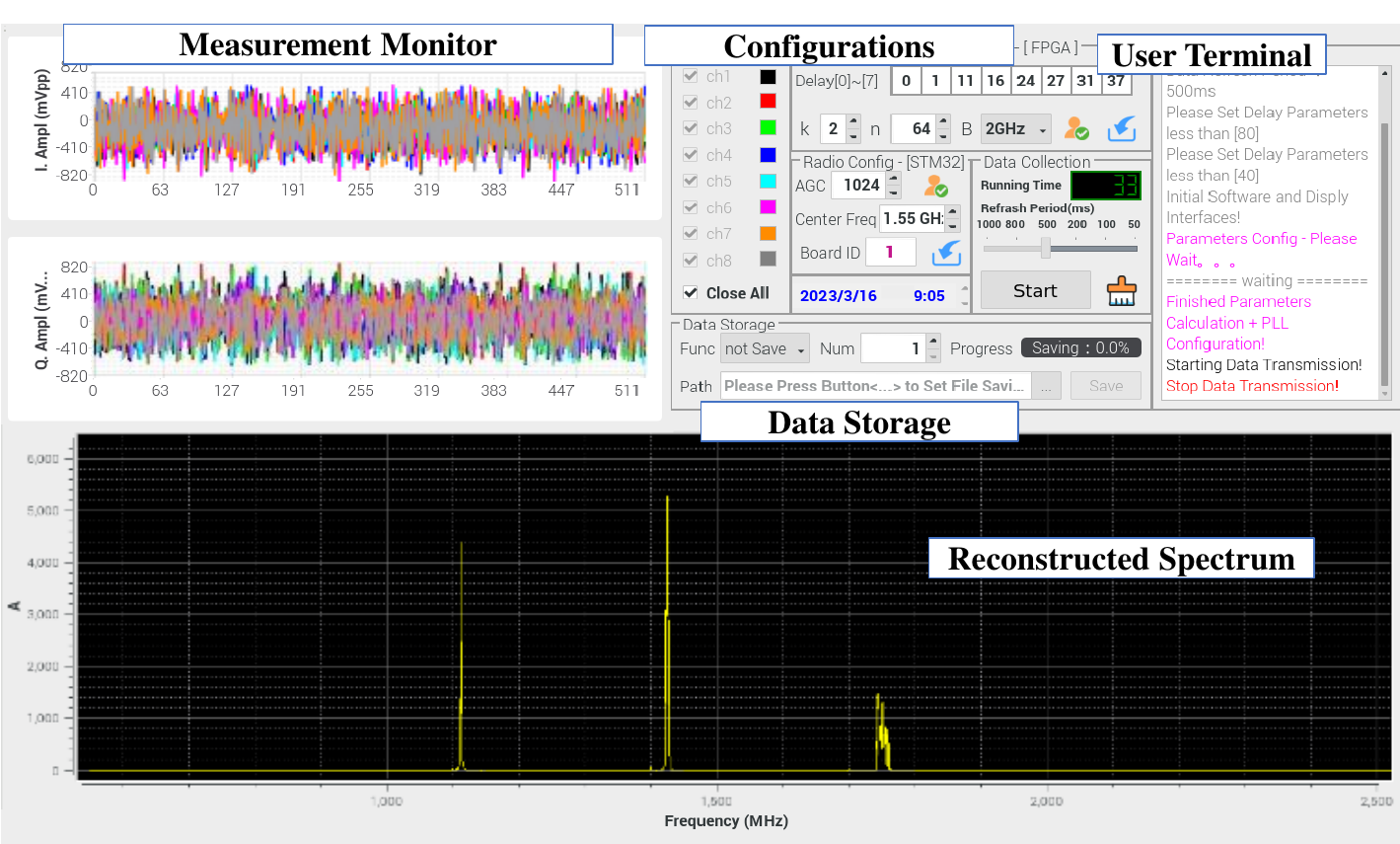}\label{fig_panel}}
\subfigure[]{\includegraphics[width=1\linewidth]{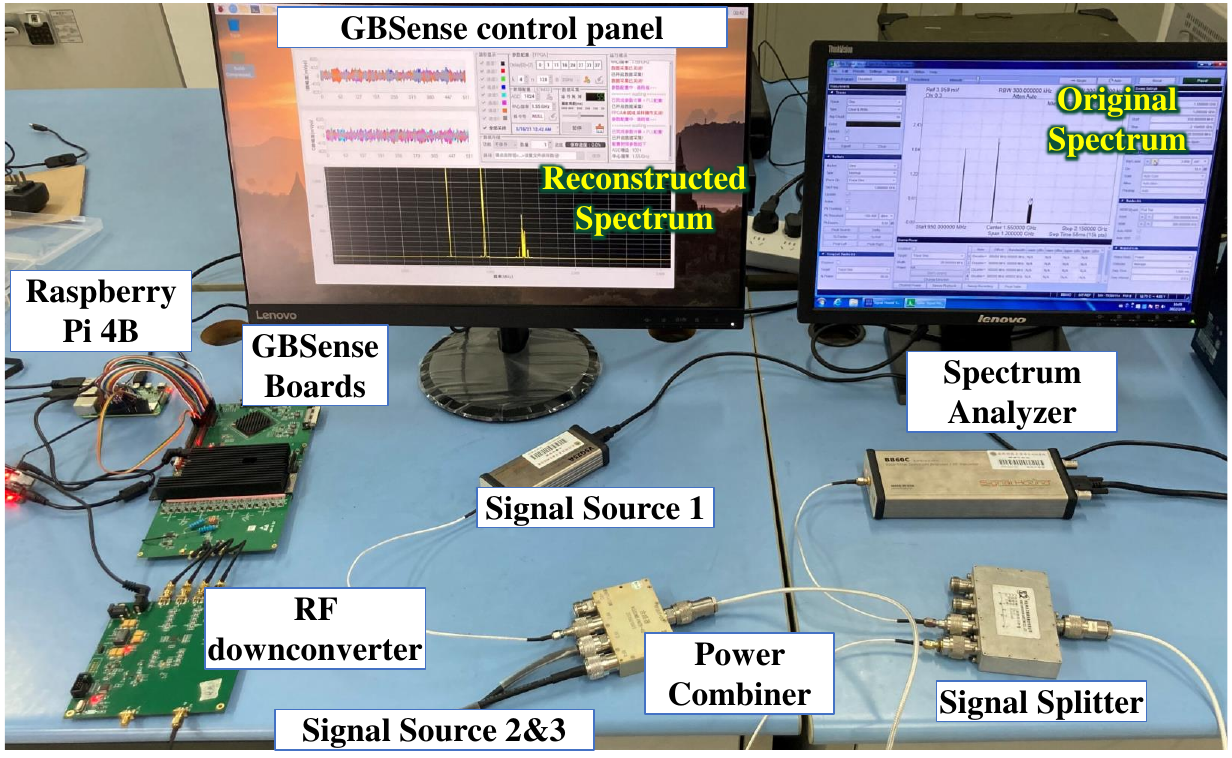}\label{fig_experiment}}
\caption{(a) User interface of the control and data processing software developed on the Raspberry Pi 4B, and (b) the setup of the integrated CSS system in the lab environment.}
\end{figure}

An integrated CSS system based on the GBSense sampling architecture has been successfully developed and tested. This system includes an RF downconverter board that downconverts a 950 to 2150 MHz RF signal to baseband I/Q signals with a 600 MHz single-sided bandwidth, and a low-power backend controller built on a Raspberry Pi 4B. The controller is connected to the GBSense system's LD board through low-speed GPIOs. A dedicated signal collection and processing software has been developed on the Raspberry Pi, incorporating embedded spectrum reconstruction algorithms. The software supports real-time system configuration, snapshot-based spectrum reconstruction, and raw PNS sampling data storage when preprocessing (data decimation and FFT) on the LD is switched off, as shown in the control panel (Fig. \ref{fig_panel}).

The effectiveness of the integrated CSS system is demonstrated in a laboratory environment, as shown in Fig. \ref{fig_experiment}. Three signals are individually generated within the 950 to 2150 MHz range using RF signal generators, and they are combined using a power combiner. The combined signal is then split, with one feed directed into the CSS system and the other analyzed directly by a spectrum analyzer for ground truth comparison. The CSS system shows consistent performance with the ground truth, verifying the accuracy and reliability of the spectrum reconstruction.

Additionally, the system demonstrates impressive real-time performance, with a processing latency as low as 30 ms to process a snapshot of 1024 samples (per lane). This low latency highlights the system's capability for real-time signal processing, making it suitable for various practical applications. Overall, the successful construction and operation of the integrated CSS system showcase its effectiveness and potential for real-world deployment.

\section{Conclusion}
This paper presents the GBSense system, a novel sub-Nyquist spectrum sensing solution for wideband signals with bandwidths up to 2 GHz. By leveraging periodic nonuniform sampling and time-interleaved analog-to-digital conversion, GBSense significantly reduces the required sampling rate to an average of 400 Msps while maintaining accurate spectrum reconstruction. Through lab experiments, the system demonstrated 100\% confidence in spectrum occupancy detection below 100 MHz and over 80\% confidence for spectrum occupancy up to 200 MHz. Additionally, with an integrated compressed spectrum sensing system built around a low-power Raspberry Pi processor, the system achieves a low processing latency of approximately 30 ms per frame, enabling real-time performance. The hardware implementation, including an optimized clock distribution circuit and commercially available components, allows for flexible adjustment of sampling parameters without requiring complex analog delays. The GBSense system provides a hardware-friendly and cost-effective solution for wideband spectrum sensing in future dynamic spectrum access systems, with potential applications in 5G, 6G, and beyond. 

 \bibliographystyle{IEEEtran}
\bibliography{reference}

\appendices

\section{Formulation of the CSS Problem}

By fixing a sampling window of length $N$, the MCS measurement is denoted as $P$ row vectors: 
\begin{equation}\label{eq_mcasample}
    \mathbf{x}_{c_p} = \left[\;x_{c_p}[1]\;x_{c_p}[2]\;\cdots\;x_{c_p}[N]\;\right], \quad p=1,2,...,P.
\end{equation}

We denote a uniform sampled $x(t)$ at rate $Lf_{\mathrm{s}}$ as $x[n]=x(t+n\frac{1}{Lf_{\mathrm{s}}})=x(t+n\frac{T}{L})$, and Denote the $NP$-point FFT of $x[n]$ as the original spectrum $\theta[n]$
as
The reconstruction of  $\theta[n]$ from the samples \eqref{eq_mcasample} can be formulated as a multiple measurement vectors (MMV) problem:
\begin{equation}
\mathbf{X} = \mathbf{A\Theta},
\end{equation}
where $\mathbf{X}\in\mathbb{C}^{P\times N}$ is the measurement, whose $(p,n)$-th element is:
\begin{equation}
X_{p,n}={Te^{-j2\pi  c_pn/(LN)}}
{\sum_{i=1}^{N}x_{c_p}[i]e^{-j2\pi  ni/N}},
\end{equation}
and $\mathbf{A}\in\mathbb{P}^{R\times L}$ is the sensing matrix with
\begin{equation}\label{eq_A}
{A}_{p,l} = e^{ j2\pi c_p\cfrac{(l-1)}{L}},
\end{equation}
and each row of $\mathbf{\Theta}\in\mathbb{C}^{L\times N}$ corresponds to the sections of the $LN$-points original spectrum $\theta[n]$ with 
\begin{equation}
\mathbf{\Theta}_{l,n} = \theta[lL+n].
\end{equation}

\section{Mathematical Equivalence Between Traditional MCS and GBSense Sampling Architectures}
Here, we demonstrate the frequency-domain equivalence between the traditional MCS and the GBSense sampling architectures, since the precise timing of the samples is less important in CSS systems.
The traditional MCS structure in Fig. \ref{fig:mcs_traditional} shows an aligned parallel sampling after an analog delay filter array. For the $p$-th lane, the delayed analog signal is expressed as $x_{c_p}(t)=x(t+\frac{c_p}{L}T)$. The behavior of the ADC that works at rate $f_{\mathrm{s}}=1/T$ can be described by a Dirac comb:
\begin{equation}
    \Sh_{T}(t) \coloneqq \sum_{k=-\infty}^{\infty}\delta (t-kT).
\end{equation}
Thus, the discrete-time Fourier transform (DTFT) of time-domian samples is calculated by
\begin{equation}\label{eq_mcs}
\begin{aligned}
     &\quad X_{c_p}^{\mathrm{MCS}}\left(e^{j2\pi fT}\right) \\
     &= \mathcal{F}\left\{x(t+\frac{c_p}{L}T)\cdot\Sh_{T}(t)\right\}\\
     &=\left[\int_{-\infty}^{+\infty}x(t+\frac{c_p}{L}T)e^{-j2\pi ft}dt\right] * \left[\frac{1}{T}\Sh_{\frac{1}{T}}(f)\right]\\
     &=\left[\int_{-\infty}^{+\infty}x(t)e^{-j2\pi ft+j2\pi fc_pT/L}dt\right] * \left[\frac{1}{T}\Sh_{\frac{1}{T}}(f)\right]\\
     &=\left[X(f)e^{j2\pi fc_pT/L}\right] * \left[\frac{1}{T}\Sh_{\frac{1}{T}}(f)\right]\\
     &=\frac{1}{T}e^{j2\pi c_pTf/L}\sum_{n=-\infty}^{+\infty}X(f-\frac{n}{T})e^{-j2\pi\frac{nc_p}{L}}.
\end{aligned}
\end{equation}
The GBSense sampling architecture in Fig. \ref{fig:mcs_gbsense} shows a non-uniform sampling followed by a re-align operation. This sampling process is denoted by process B. For the $p$-th lane, the undelayed signal $x(t)$ is sampled by $\Sh_{T}(t-\frac{c_p}{L}T)$. The result is further realigned by $\delta(t-\frac{c_p}{L}T)$. The DTFT of the samples is calculated by
 \begin{equation}\label{eq_gbsense}
\begin{aligned}
     &\quad X_{c_p}^{\text{GBSense}}\left(e^{j2\pi fT}\right) \\
     &= \mathcal{F}\left\{\left[x(t)\cdot\Sh_{T}(t-c_iT/L)\right]*\delta(t-\frac{c_p}{L}T)\right\}\\
     &=e^{j2\pi c_pTf/L} \cdot \left[X(f) * \left[\frac{1}{T}\cdot e^{-j2\pi f c_pT/L}\cdot\Sh_{\frac{1}{T}}(f)\right]\right]\\
     &=\frac{1}{T}e^{j2\pi c_pTf/L}\int_{-\infty}^{+\infty}X(\zeta)\cdot \left[e^{-j2\pi (f-\zeta) c_iT/L}\cdot\Sh_{\frac{1}{T}}(f-\zeta)\right]\mathrm{d}\zeta\\
     &=\frac{1}{T}e^{j2\pi c_pTf/L}\sum_{n=-\infty}^{+\infty}X(f-\frac{n}{T})e^{-j2\pi\frac{nc_p}{L}}
\end{aligned}
\end{equation}
We have demonstrated the equivalence between \eqref{eq_mcs} and \eqref{eq_gbsense}, which indicates that signal reconstruction algorithms designed for traditional MCS can be directly applied to GBSense samples.

\vfill

\end{document}